\documentclass[10pt, conference, letterpaper]{IEEEtran}
\IEEEoverridecommandlockouts
\usepackage{cite}
\usepackage{amsmath,amssymb,amsfonts}
\usepackage{algorithmic}
\usepackage{graphicx}
\usepackage{textcomp}
\usepackage{xcolor}

\usepackage{algorithm,algorithmic,mathtools}

\usepackage{multirow}

\usepackage{subfigure}
\usepackage{pgfplots}
\usetikzlibrary{pgfplots.groupplots}
\pgfplotsset{compat=1.3}

\def\BibTeX{{\rm B\kern-.05em{\sc i\kern-.025em b}\kern-.08em
    T\kern-.1667em\lower.7ex\hbox{E}\kern-.125emX}}
\begin{document}

\title{Virtual Network Embedding\\without Explicit Virtual Network Specification}

\author{\IEEEauthorblockN{Jiangnan Cheng, Yingjie Bi, and Ao Tang}
\IEEEauthorblockA{
School of Electrical and Computer Engineering, Cornell University\\
Ithaca, NY, USA \\
\{jc3377, yb236, atang\}@cornell.edu}
}

\maketitle

\begin{abstract}
Network virtualization enables Internet service providers to run multiple heterogeneous and dedicated network architectures for different customers on a shared substrate. In existing works on virtual network embedding (VNE), each customer formulates a virtual network request (VNR) where a virtual network (VN) is required. Motivated by a concrete example where VN is not a proper VNR formulation to reflect the traffic demand of a customer, we propose a new VNR formulation described by the traffic demand between several access node pairs to complement the existing VNR formulation. Moreover, three different groups of VNE variants
% , which together form six different embedding settings, 
are systematically examined. Simulations demonstrate that shared channel embedding, as a new embedding variant under the proposed VNR formulation, improves the acceptance rate and reduces cost and link utility compared to traditional independent channel embedding.

% It requires efficient techniques for virtual network embedding (VNE) -- mapping virtual nodes and links to the corresponding physical ones in the substrate network (SN).  

% When a customer is allowed to specify a joint bandwidth capacity for a group of links in virtual network request (VNR), which is usually smaller than the sum of their individual bandwidth capacities, VNE can potentially generate a smaller and hence cheaper SN subset. 

\end{abstract}


\section{Introduction} \label{s_introduction}

The emergence of various new Internet applications demands a flexible and fast-moving network infrastructure that can handle increasingly larger range of traffic dynamics, satisfy the diverse real-time characteristics and rapidly deliver novel network service to customers. On the other hand, technology development at the side of Internet service providers (ISPs) usually has a slow pace, with long verification phase and dedicated hardware to guarantee reliance and stability \cite{mijumbi2015network,han2015network}. To address this challenge, network virtualization \cite{anderson2005overcoming} is proposed to move network resources from dedicated hardware to virtualized overlays, making it possible for ISPs to develop technologies in fast cycle and run multiple heterogeneous and dedicated network architectures for different customers on a shared substrate in the meantime.

In existing works on network virtualization \cite{fischer2013virtual,yu2008rethinking, chowdhury2009virtual, chowdhury2011vineyard, cheng2011virtual, rahman2013svne, lu2006efficient, yao2018novel, yan2020automatic,zhang2023multi}, each customer formulates its traffic demand as a virtual network (VN) with a given topology and link capacities. The request of each customer is referred to as virtual network request (VNR). ISP therefore considers a virtual network embedding (VNE) problem which allocates a subset of the shared substrate network (SN) to a VN. When the SN can no longer support a VN, a fail signal should be returned to the customer. 

\begin{figure}[tbp]
\centering
\includegraphics[scale=0.5]{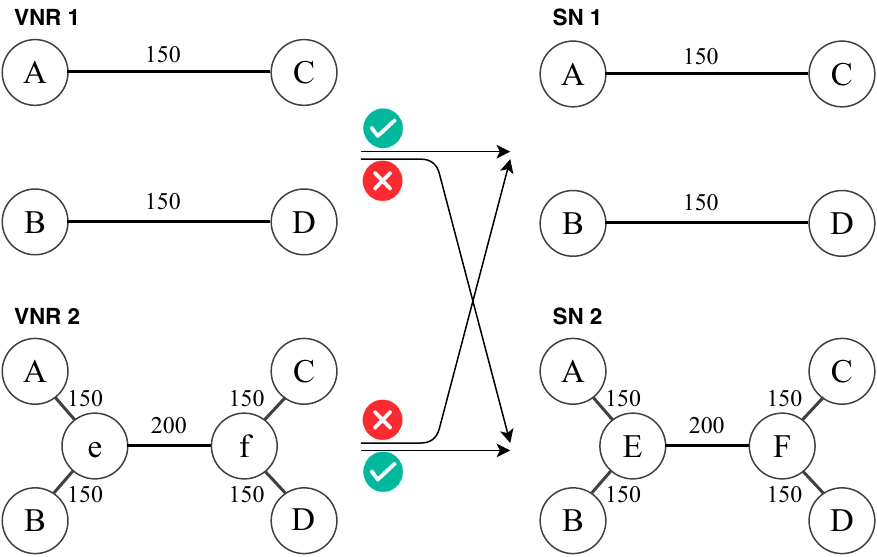}
\caption{A motivating example which illustrates the limitation of specifying VN in a VNR. VNR 1 cannot be embedded on SN 2, because each node pair can send traffic with a throughput of 150 units simultaneously in VNR 1 but not in SN 2; VNR 2 cannot be embedded on SN 1, because node A and D are connected in VNR 2 but not in SN 1. Node $e$ and $f$ in VNR 2 are virtual nodes and can be mapped to arbitrary SN nodes.}
\label{fig_motivation}
\end{figure}

Obviously, the most economical way to formulate VNR is to specify a minimum VN which has enough link capacities to satisfy a customer's traffic demand. However, there are maybe more than one economical VNR formulations and each has its own specific limitation. 
For example, in Fig. \ref{fig_motivation}, VNR 1 and VNR 2 are two economical VNR formulations for a customer whose traffic demand for either access node pair (A,C) or (B,D) is no more than 150 units, and joint traffic demand for these two node pairs is no more than 200 units. Intuitively, VNR 2 is more economical than VNR 1 since the former one characterizes the joint traffic demand through a VN link (e,f) while the latter one does not. As a result, VNR 2 can be embedded on SN 2 while VNR 1 cannot. However, as a sacrifice, VNR 2 has to have an unnecessarily connection of node pair (A, D), preventing it to be embedded on SN 1, which on the other hand admits VNR 1. Hence, specifying VN in a VNR may not be a proper way to reflect the customer's traffic demand in some cases.

Therefore, in this paper, we propose a VNR formulation which directly characterizes the customer's traffic demand for a group of access node pairs, without specifying a particular virtual network. Moreover, under the new VNR formulation, we are able to consider three different groups of VNE variants, including single-path vs multi-path, channel vs shared channel, and oblivious routing vs adaptive routing (both belong to shared channel). Among them, independent channel embedding can be reduced to existing VNE problems in previous works \cite{yu2008rethinking, chowdhury2009virtual, chowdhury2011vineyard, cheng2011virtual, rahman2013svne, lu2006efficient, yao2018novel, yan2020automatic,zhang2023multi}. Starting from the well-known approaches for independent channel embedding, we propose new algorithms to address shared channel embedding.

The rest of the paper is organized as follows. 
% In Section \ref{s_motivation}, we present a motivating example in which VN topology limits the search space for valid SN subset. 
In Section \ref{s_formulation}, we formulate the basic problem of VNE, and then discuss three different groups of VNE variants. In Section \ref{s_ufp_embedding} and Section \ref{s_mcf_embedding}, we present algorithms for single-path embedding and multi-path embedding, respectively. The performances of different embedding algorithms are compared in Section \ref{s_evaluation}. 
Section \ref{s_discussion} discusses the scenarios where node constraints and delay constraints are considered and link cost is a piecewise linear or quadratic function of link bandwidth, 
and Section \ref{s_conclusion} concludes the paper.

\section{Problem Formulation} \label{s_formulation}

In this section, we first describe the basic VNE problem where VNR directly characterizes the customer's traffic demand. We then introduce three different groups of VNE variants.
% which together can form six different embedding settings. 

% In the end, we discuss the potential problem of single-path oblivious routing embedding due to which it is no longer considered in the following sections.

\subsection{Basic VNE Problem} \label{ss_basic}

In most VNE researches, each virtual node is allowed to be mapped to different real nodes of SN, and therefore corresponding VNE is decomposed into two phases: node mapping and link mapping. However, here we only consider the most basic VNE problem where each node appeared in VNR are real nodes of SN. This is equivalent to the scenario where virtual node locations are predetermined, and we adopt this configuration since our main focus is formulate the joint traffic demand constraints in the VNE, which only modifies the link bandwidth constraints and thus mainly affects link mapping.
The discussion of node constraints, as an independent topic, is postponed to Section \ref{s_discussion}.

\textbf{SN.} Suppose the SN is represented by a weighted undirected graph $G^s = (V^s, E^s)$, where $V^s$ and $E^s$ denote the set of nodes and links, respectively. The link that connects node $i$ and $j$ (both $\in V^s$) is denoted by a two-field tuple $(i,j)$. 
% $\forall e^s\in E^s$, 
The available bandwidth and unit price of each link $e^s \in E^s$ are denoted by $c^s(e^s)$ and $p^s(e^s)$, respectively.

\textbf{VNR.} Each VNR requires an SN subset that has the ability to deliver traffic between $N$ access node pairs given by node pair vector $\mathbf{e}^v = ((s_1, t_1), (s_2, t_2), \cdots, (s_N, t_N))^\top$, where $s_n, t_n \in V^s$, $\forall n \in \{1, 2, \cdots, N\}$. The $n$-th element of $\mathbf{e}^v$ is denoted by $\mathbf{e}^v(n)$. The associated traffic demand of $\mathbf{e}^v$ is denoted by traffic demand vector $\mathbf{d}^v \in \mathbb{R}^{N \times 1}$, the $n$-th element of whom, denoted by $\mathbf{d}^v(n)$, represents the traffic demand for $\mathbf{e}^v(n)$. Suppose the upper bound of $\mathbf{d}^v$ is described by matrix inequality $\mathbf{A}^v\mathbf{d}^v \leq \mathbf{b}^v$, where $\mathbf{A}^v \in \mathbb{R}^{M^v\times N}_{\geq 0}$ and $\mathbf{b}^v \in \mathbb{R}^{M^v\times 1}_{+}$; and the lower bound of $\mathbf{d}^v$ is described by $\mathbf{d}^v \geq \mathbf{0}$. For convenience we denote $\{\mathbf{d}^v|\mathbf{A}^v\mathbf{d}^v \leq \mathbf{b}^v, \mathbf{d}^v \geq \mathbf{0}\}$, the feasible region of $\mathbf{d}^v$ with respect to this VNR, by $D^v$; and we also define $\mathbf{d}^v_{\text{max}}(n) = \max \{\mathbf{d}^v(n) | \mathbf{d}^v \in D^v\}$, which represents the maximum possible traffic demand for $\mathbf{e}^v(n)$ with respect to this VNR. The VNR can therefore be denoted by a three-field tuple $R^v = (\mathbf{e}^v, \mathbf{A}^v, \mathbf{b}^v)$.
% Our formulation of VNR describes the traffic demand in a finer way because the joint traffic demand, if any, is also considered.

The key advantage of such VNR formulation is its ability to properly characterize the joint traffic demand upper bounds. For example, the VNR described in Section \ref{s_introduction} can be formulated as
\begin{align}
\mathbf{e}^v & = ((\text{A}, \text{C}), (\text{B}, \text{D}))^\top, \label{eq_req_example_start}\\
\mathbf{A}^v & = \begin{bmatrix}
1 & 0 & 1\\
0 & 1 & 1
\end{bmatrix}^\top,\\
\mathbf{b}^v & = [150, 150, 200]^\top. \label{eq_req_example_end}
\end{align}
% and the upper bound of $\mathbf{d}^v$ is described by
% \begin{align}
% \begin{bmatrix}
% 1 & 0 \\
% 0 & 1 \\
% 1 & 1
% \end{bmatrix}
% \mathbf{d}^v
% \leq
% \begin{bmatrix}
% 150 \\
% 150 \\
% 200
% \end{bmatrix}.
% \label{eq_req_example}
% \end{align}
% $\mathbf{A}^v = \begin{bmatrix}
% 1 & 0 & 1\\
% 0 & 1 & 1
% \end{bmatrix}^\top$, and $\mathbf{b}^v = [150, 150, 200]^\top$.

\textbf{VNE.} For a given VNR $R^v$, VNE allocates a subset of SN that can satisfy any valid traffic demand of $R^v$ (a fail signal is returned if there's no feasible allocation). The topology of the subset is denoted by $G^v = (V^v, E^v)$, and the allocated bandwidth for each link $e^v \in E^v$ is $c^v(e^v)$.
% Moreover, we denote the routing solution associated with $G^r$ for each $\mathbf{d}^v \in D^v$ by $\mathbf{\Omega}(G^r, \mathbf{d}^v) \in \mathbb{R}^{|E^r|\times N}$. $\forall i, j$, $\mathbf{\Omega}_{ij}(G^r, \mathbf{d}^v)$ specifies the amount of $\mathbf{d}^v(j)$ that should be sent on the $i$-th link of $E^r$. 
For convenience of computation, in the rest of the paper, we let $V^v = V^s$ and $E^v = E^s$, and allow $c^v(e^s)$ to be zero for some link $e^s \in E^s$.

\textbf{Objective.} The objective of VNE is to find the bandwidth allocation $c^v$ that minimizes the cost:
\begin{align}
\min_{c^v(e^s)} \quad \sum_{e^s\in E^s} p^s(e^s)c^v(e^s). \label{eq_objective}
\end{align}

\subsection{VNE Variants} \label{ss_req}

\begin{figure*}[tbp]
\centering
\subfigure[SN 3\label{fig_ufp_sn}]{
    \includegraphics[scale=0.70]{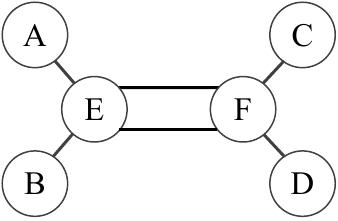}}
\hfill
\subfigure[Single-path, independent channel\label{fig_ufp_normal}]{
    \includegraphics[scale=0.70]{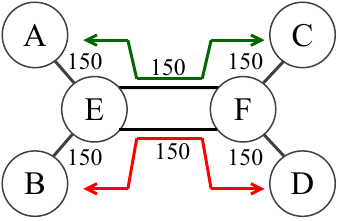}}
\hfill
\subfigure[Single-path, oblivious routing\label{fig_ufp_oblivious}]{
    \includegraphics[scale=0.70]{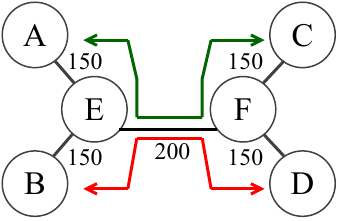}}
\hfill
\subfigure[Single-path, adaptive routing\label{fig_ufp_adaptive}]{
\includegraphics[scale=0.70]{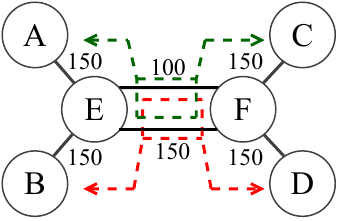}}

\subfigure[SN 4\label{fig_mcf_sn}]{
    \includegraphics[scale=0.70]{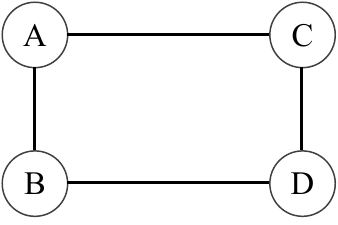}}
\hfill
\subfigure[Multi-path, independent channel\label{fig_mcf_normal}]{
    \includegraphics[scale=0.70]{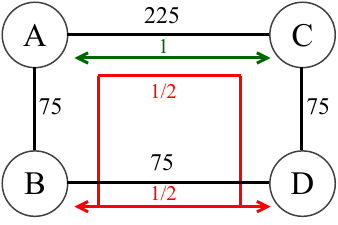}}
\hfill
\subfigure[Multi-path, oblivious routing\label{fig_mcf_oblivious}]{
    \includegraphics[scale=0.70]{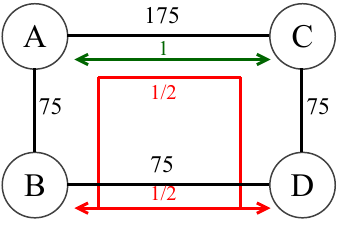}}
\hfill
\subfigure[Multi-path, adaptive routing\label{fig_mcf_adaptive}]{
\includegraphics[scale=0.70]{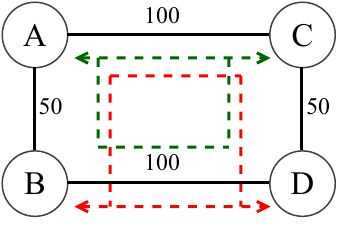}}
\caption{VNE with different requirements}
\label{fig_vne_req}
\end{figure*}

VNE can have different variants that impose additional constraints on the allocation. 
% The quality of resulted $G^r$, such as whether the in-order delivery of data packets is preserved, whether a drastically changing traffic demand can be accommodated, etc., can also be different. 
In this paper, we consider three groups of VNE variants as follows, which together can form six different embedding settings.

\textbf{Single-path VS Multi-path.} Single-path embedding requires the traffic of each node pair to be routed on only one path, while for multi-path embedding the traffic of each node pair can be routed on multiple paths simultaneously. Multi-path embedding is more flexible, but it can lead to out-of-order delivery problem of data packets \cite{yu2008rethinking}.  

\textbf{Independent Channel VS Shared Channel.} In independent channel embedding, an independent channel (either single-path or multi-path) with end-to-end bandwidth $\mathbf{d}^v_{\text{max}}(n)$ has to be offered to node pair $n$, $\forall n$. So in VNR, $\mathbf{A}^v$ and $\mathbf{b}^v$ can be equivalently replaced by $I_n$ (i.e., $n \times n$ identity matrix) and $\mathbf{d}^v_{\text{max}}$, respectively. Hence independent channel embedding can be reduced to those VNE problems in previous works \cite{yu2008rethinking, chowdhury2009virtual, chowdhury2011vineyard, cheng2011virtual, rahman2013svne, lu2006efficient, yao2018novel, yan2020automatic,zhang2023multi} by constructing a VN where link $\mathbf{e}^v(n)$ has bandwidth $\mathbf{d}^v_{\text{max}}(n)$, $\forall n$. On the other hand, in shared channel embedding, different node pairs are allowed to share the bandwidth of a common link, which generally results in a cheaper allocation by considering the joint traffic demand upper bounds. 
% However, shared channel embedding problem is at least no easier than independent channel embedding problem, because the latter one can be reduced to the former one by just specifying the individual traffic demand of each node pair, as discussed in Section \ref{ss_basic}.  

\textbf{Oblivious Routing VS Adaptive Routing.} There are two different routing strategies in shared channel embedding: oblivious routing and adaptive routing \cite{azar2004optimal, kinsy2009application}. In oblivious routing, the traffic between a node pair is routed on one or more predefined paths with predefined path splitting ratios. On the other hand, in adaptive routing, the paths and the associated path splitting ratios for the traffic between a node pair are dynamically adjusted according to real-time network congestion. Adaptive routing generally results in a cheaper allocation by offering more flexibility, yet the performance is affected by routing speed and complexity \cite{kinsy2009application}, so it may not be able to accommodate a drastically changing traffic demand.

Two examples that demonstrate the differences of VNE variants ares shown in Fig. \ref{fig_vne_req}. In these two examples, we consider the same VNR described in Equation (\ref{eq_req_example_start})-(\ref{eq_req_example_end}).

% which $\mathbf{e}^v = ((\text{A}, \text{C}), (\text{B}, \text{D}))^\top$, and the upper bound of $\mathbf{d}^v$ is described by
% \begin{align}
% \begin{bmatrix}
% 1 & 0 \\
% 0 & 1 \\
% 1 & 1
% \end{bmatrix}
% \mathbf{d}^v
% \leq
% \begin{bmatrix}
% 150 \\
% 150 \\
% 200
% \end{bmatrix}
% \label{eq_req_example}
% \end{align}

In the first example, we consider single-path embedding. Fig. \ref{fig_ufp_sn} shows the SN topology. Fig. \ref{fig_ufp_normal}, Fig. \ref{fig_ufp_oblivious} and Fig. \ref{fig_ufp_adaptive} show feasible $G^v$ examples for independent channel, oblivious routing and adaptive routing embedding, respectively. It is obvious that the $G^v$'s in Fig. \ref{fig_ufp_normal} and Fig. \ref{fig_ufp_oblivious} can satisfy any valid traffic demand in their corresponding ways. And in Fig. \ref{fig_ufp_adaptive}, we always route the smaller and larger one among $\mathbf{d}^v(1)$ and $\mathbf{d}^v(2)$ on the upper and lower link between E and F, respectively. When the smaller and larger one among $\mathbf{d}^v(1)$ and $\mathbf{d}^v(2)$ exchange, the path for each node pair will switch with each other, but remain to be a single path. 
% According to Equation (\ref{eq_req_example}), in this way the bandwidths of the two links between E and F are just enough.

In the second example, we consider multi-path embedding. Fig. \ref{fig_mcf_sn} shows the SN topology. Fig. \ref{fig_mcf_normal}, Fig. \ref{fig_mcf_oblivious} and Fig. \ref{fig_mcf_adaptive} show feasible $G^v$ examples for independent channel, oblivious routing and adaptive routing embedding, respectively. It is obvious that the $G^v$'s in Fig. \ref{fig_mcf_normal} and Fig. \ref{fig_mcf_oblivious} can satisfy any valid traffic demand in their corresponding ways. And in Fig. \ref{fig_mcf_adaptive}, for each node pair we route the its traffic on the link that directly connects it first. When the amount of its traffic is larger than 100 units, the exceeding part is routed through the other two nodes. 
% According to Equation (\ref{eq_req_example}), at most one node pair can have traffic more than 100 units, and the exceeding part can at most be 50 units, therefore the bandwidths of all the links are just enough.

Since single-path adaptive routing has the problem of shifting paths back and forth under small disturbance, we don't consider it in the following sections.

% Therefore, in this paper we do not consider single-path adaptive routing which may shift path back and forth under small disturbance.    

\section{Single-Path VNE Algorithm} \label{s_ufp_embedding}

Single-path independent channel (SPIC) embedding problem can be reduced to Unsplittable Flow Problem (UFP), which is NP-hard \cite{kleinberg1996approximation}. Since independent channel embedding can be viewed as a special case of shared channel embedding with $\mathbf{A}^v = I_n$, single-path oblivious routing (SPOR) embedding problem is also NP-hard. Therefore, approximation approaches are generally adopted to solve single-path embedding problems.
In this section, we present the approximation algorithms for solving SPIC and SPOR embedding problem in Section \ref{ss_ufp_independent} and \ref{ss_ufp_oblivious}, respectively.

\subsection{Single-path Independent Channel (SPIC)} \label{ss_ufp_independent}

\begin{algorithm}[tp]
    \caption{SPIC algorithm}
    \begin{algorithmic}[1]
    \STATE Initialize $c^v(e^s) = 0$, $\forall e^s \in E^s$
    \STATE Compute $\mathbf{d}^v_{\text{max}}$ through LP

    \FOR{each $n \in \{1, 2, \cdots, N\}$}
        \STATE Search the $K$-least cost paths on the SN between node pair $\mathbf{e}^v(n)$
        \FOR{$k \in \{1, 2, \cdots, K\}$}
            \FOR{each link $e^s$ along path $k$}
                \STATE Assume node pair $\mathbf{e}^v(n)$ uses path $k$, and compute the target bandwidth allocation for link $e^s$
                \begin{align}
                \hat{c}^v(e^s) := c^v(e^s) + \mathbf{d}^v_{\text{max}}(n) \label{eq_ufp_ind_link_capacity}
                \end{align}
                \STATE If $\hat{c}^v(e^s) > c^s(e^s)$, \textbf{break} 
            \ENDFOR 
            \STATE If $\hat{c}^v(e^s) \leq c^s(e^s)$ for each link $e^s$ along path $k$, update $c^v(e^s) \leftarrow \hat{c}^v(e^s)$, set path $k$ as the path for node pair $\mathbf{e}^v(n)$, and \textbf{break}; or else \textbf{continue} 
        \ENDFOR
        \STATE If no path is set for node pair $\mathbf{e}^v(n)$, \textbf{return} fail signal 
    \ENDFOR
    
    \STATE Embedding succeeds, \textbf{return} the allocation $c^v$
    \end{algorithmic}
    \label{alg_ufp_independent}
\end{algorithm}

The algorithm for SPIC embedding is shown in Algorithm \ref{alg_ufp_independent}, which was proposed in \cite{yu2008rethinking}. In line 1, we first initialize $c^v(e^s) = 0$, $\forall e^s \in E^s$. In line 2, we then compute $\mathbf{d}^v_{\text{max}}$. This can be done by solving $N$ linear programming (LP) problems, which takes polynomial time. Next, in line 3-13, for each $n \in \{1, 2, \cdots, N\}$, we search the $K$-least unit cost paths
% \footnote{Here the cost of each link $e^s$ is set to $c^s(e^s)\mathbf{d}^v_{max}(n)$ and the cost of a path equals the sum of costs of all the links belong to it. Since here we have fixed unit price for each link, we also can compute $K$-least unit price paths, which will lead to the same result.}
on the SN between node pair $\mathbf{e}^v(n)$ for a fixed $K$, which can be done in $O(|E^v| + |V^v|\log|V^v| + K)$ time \cite{eppstein1998finding}. For each potential path $k$, we assume it is used by node pair $\mathbf{e}^v(n)$ and check whether we can allocate enough bandwidths for all the links along it. For each link $e^s$ along path $k$, its target bandwidth allocation, denoted by $\hat{c}^v(e^s)$, is given by Equation (\ref{eq_ufp_ind_link_capacity}). Once we have $\hat{c}^v(e^s) \leq c^s(e^s)$ for each link $e^s$ along path $k$, we set path $k$ as the path for pair $\mathbf{e}^v(n)$. If no path is set, a fail signal indicating the failure of VNE should be returned. In the end, in line 14, if for each node pair we successfully set a path, the embedding succeeds and the resulted allocation $c^v$ is returned.

% to pass EDAS check
\addtolength{\topmargin}{0.01in}

\subsection{Single-path Oblivious Routing (SPOR)} \label{ss_ufp_oblivious}

Our algorithm for SPOR embedding differs from Algorithm \ref{alg_ufp_independent} by replacing Equation (\ref{eq_ufp_ind_link_capacity}) with

\begin{align}
\hat{c}^v(e^s) :=
\begin{bmatrix}
\underset{\mathbf{d}^v}{\text{max}} \quad & \underset{n' \leq n:\;\text{pair }n'\text{ uses } e^s}{\sum}  \mathbf{d}^v(n') \\
s.t. \quad & \mathbf{A}^v\mathbf{d}^v \leq \mathbf{b}^v, \mathbf{d}^v \geq \mathbf{0}
\end{bmatrix} \label{eq_ufp_oblivious_link_capacity}      
\end{align}
and removing line 2 of Algorithm \ref{alg_ufp_independent}. Equation (\ref{eq_ufp_oblivious_link_capacity}) dictates that the target bandwidth allocation $\hat{c}^v(e^s)$ should be the maximum of the sum of $\mathbf{d}^v(n')$'s, where $n'$ can be the index of any node pair that uses $e^s$ to deliver traffic, i.e., $e^s$ is on its path. Such $\hat{c}^v(e^s)$ can be computed through LP. It is no larger than $c^v(e^s) + \mathbf{d}^v_{\text{max}}(n)$
% , because the amount of traffic for node pair $\mathbf{e}^v(n)$ is no larger than $\mathbf{d}^v_{\text{max}}(n)$; 
% Moreover, $\hat{c}^r(e^s)$ can be potentially smaller than $c^r(e^s) + \mathbf{d}^v_{\text{max}}(n)$
and can be potentially smaller, because node pair $\mathbf{e}^v(n)$ may share $e^s$ with other node pairs and they may have a joint traffic demand upper bound smaller than the sum of their individual traffic demand upper bounds. Therefore, our SPOR algorithm generally outperforms Algorithm \ref{alg_ufp_independent}.

\section{Multi-Path VNE Algorithm} \label{s_mcf_embedding}

Multi-path independent channel embedding (MPIC) problem can be reduced to Multi-commodity Flow Problem (MFP) \cite{assad1978multicommodity}, which is solvable in polynomial time. On the other hand, multi-path shared channel embedding problem may not necessarily be polynomial. In this paper, we find a polynomial time algorithm to solve multi-path oblivious routing (MPOR) embedding problem, while the complexity of multi-path adaptive routing (MPAR) embedding problem remains unclear.

In this section, we present the algorithms for solving MPIC, MPOR and MPAR embedding problem in Section \ref{ss_mcf_independent}, \ref{ss_mcf_oblivious} and \ref{ss_mcf_adaptive}, respectively.

\subsection{Multi-path Independent Channel (MPIC)} \label{ss_mcf_independent}

MPIC embedding problem can be reduced to MFP and it is solvable in polynomial time. Similar to SPIC embedding in Section \ref{ss_ufp_independent}, we first compute $\mathbf{d}^v_{\text{max}}$ in polynomial time. Next, $\forall n \in \{1, 2, \cdots, N\}, \forall e^s \in E^s$, we define two flow variables $f^v_{n}(e^s_+)$ and $f^v_{n}(e^s_-)$, which correspond to the portion of $\mathbf{d}^v(n)$ send on edge $e^s$ in two different directions. Then we can formulate the corresponding MFP in the form of LP with objective function in Equation (\ref{eq_objective}) and following constraints:

\noindent1) \textbf{Flow conservation}
\begin{align}
& \forall i \in V^s, \forall n \in \{1, 2, \cdots, N\}: \notag\\
& \quad\sum_{e^s:\exists j\text{ s.t. }e^s = (i,j)} f^v_{n}(e^s_+) -  \sum_{e^s:\exists j\text{ s.t. }e^s = (j,i)} f^v_{n}(e^s_-)\notag\\
& \quad= \left\{
            \begin{array}{ll}
              1 \; & i = s_n \\
              -1 \; & i = t_n \\
              0 \; & \text{o.w.}
            \end{array}
\right.\label{eq_ind_flow_conservation}
\end{align}

\noindent2) \textbf{Flow non-negativity}
\begin{align}
\forall n \in \{1, 2, \cdots, N\}, \forall e^s \in E^s: \quad f^v_{n}(e^s_+), f^v_{n}(e^s_-) \geq 0\label{eq_ind_flow_nonnegativity}
\end{align}

\noindent3) \textbf{SN link bandwidth}
\begin{align}
\forall e^s \in E^s: \quad c^v(e^s) \leq c^s(e^s)  \label{eq_ind_link_capacity}
\end{align}

\noindent4) \textbf{Traffic demand}
\begin{align}
& \forall e^s \in E^s:\;
({\mathbf{d}^v_{\text{max}}})^\top \mathbf{f}^v(e^s)  \leq c^v(e^s)  \label{eq_ind_link_bandwidth}
\end{align}
where $\mathbf{f}^v(e^s) \triangleq [f^v_1(e^s_+) + f^v_1(e^s_-), \cdots, f^v_N(e^s_+) + f^v_N(e^s_-)]^\top$. Here $f^v_n(e^s_+) + f^v_n(e^s_-)$ represents the sum of path splitting ratios for paths of node pair $\mathbf{e}^v(n)$ passing through link $e^s$.

\subsection{Multi-path Oblivious Routing (MPOR)} \label{ss_mcf_oblivious}

MPOR embedding problem is also solvable in polynomial time. We first formulate the problem as a robust optimization problem by replacing Equation (\ref{eq_ind_link_bandwidth}) with the following constraint in MFP in Section \ref{ss_mcf_independent}:

\begin{align}{}
& \forall e^s \in E^s:
\quad\begin{bmatrix}
    \underset{\mathbf{d}^v}{\text{max}} \quad & ({\mathbf{d}^v})^\top \mathbf{f}^v(e^s) \\
    s.t. \quad & \mathbf{A}^v\mathbf{d}^v \leq \mathbf{b}^v, \mathbf{d}^v \geq \mathbf{0} 
\end{bmatrix} \leq c^v(e^s)  \label{eq_ob_link_bandwidth}
\end{align}
which indicates that $c^v(e^s)$ should be no less than the maximum amount of traffic that sends on link $e^s$ when $\mathbf{d}^v \in D^v$.

The dual problem of the inner LP of Equation (\ref{eq_ob_link_bandwidth}) is a minimization problem:
\begin{align}
\underset{\mathbf{q}^v(e^s)}{\text{min}} \quad & (\mathbf{b}^v)^\top \mathbf{q}^v(e^s) \label{eq_ob_primal_1}\\
s.t. \quad & (\mathbf{A}^v)^\top \mathbf{q}^v(e^s) \geq \mathbf{f}^v(e^s), \label{eq_ob_primal_2}\\
& \mathbf{q}^v(e^s) \geq \mathbf{0}. \label{eq_ob_primal_3}
\end{align}
where we define dual vector $\mathbf{q}^v(e^s) \in \mathbb{R}^{M^v\times 1}$ for each link $e^s \in E^s$. Therefore, Equation (\ref{eq_ob_link_bandwidth}) can be transformed to:
\begin{align}
\forall e^s \in E^s: \quad & (\mathbf{b}^v)^\top \mathbf{q}^v(e^s) \leq c^v(e^s),\label{eq_ob_dual_1}\\
&  (\mathbf{A}^v)^\top \mathbf{q}^v(e^s) \geq \mathbf{f}^v(e^s),\label{eq_ob_dual_2} \\
& \mathbf{q}^v(e^s) \geq \mathbf{0}. \label{eq_ob_dual_3}
\end{align}
Thus the oblivious routing problem is formulated as an LP problem with objective function in Equation (\ref{eq_objective}), and with constraints in Equation (\ref{eq_ind_flow_conservation})-(\ref{eq_ind_link_capacity}) and (\ref{eq_ob_dual_1})-(\ref{eq_ob_dual_3}).

\begin{table}[tbp]
\caption{Comparison of LP input complexity}
\begin{center}
\begin{tabular}{|c|c|c|}
\hline
 & \textbf{\textit{Number of}}& \textbf{\textit{Number of }} \\
VNE & \textbf{\textit{decision variables}}& \textbf{\textit{constraints} }\\
\hline
MPIC & $O(N|E^s|)$ & $O(N|V^s| + N|E^s|)$ \\
\hline
MPOR & $O((N + M^v)|E^s|)$ & $O(N|V^s| + (N + M^v)|E^s|)$ \\
\hline
MPAR & $O(QN|E^s|)$ & $O(QN|V^s| + QN|E^s|)$ \\
\hline
\end{tabular}
\label{tab_lp_complexity}
\end{center}
\end{table}

\begin{algorithm}[tp]
    \caption{MPOR-Fast Algorithm}
    \begin{algorithmic}[1]

    \STATE Solve MPAR embedding problem
    \IF{MPAR fails}
        \STATE \textbf{return} fail signal
    \ELSE
        \STATE get the allocation $c^v$ from MPAR, and fix the corresponding path splitting ratios $\mathbf{f}^v(e^s)$
        \FOR{each $e^s \in E^s$ such that $c^v(e^s) > 0$}
            \STATE Update $c^v(e^s)$ through inner LP in Equation (\ref{eq_ob_link_bandwidth})
        \ENDFOR
    \ENDIF
    
    \STATE Embedding succeeds, \textbf{return} the allocation $c^v$
    \end{algorithmic}
    \label{alg_mcf_oblivious_aprox}
\end{algorithm}

From Table \ref{tab_lp_complexity}, we can compare the LP input complexity for MPIC and MPOR. MPOR introduces $O(M^v|E^s|)$ more decision variables and $O(M^v|E^s|)$ more constraints. which may significantly increase the running time of LP. Therefore, we can formulate an approximation algorithm as shown in Algorithm \ref{alg_mcf_oblivious_aprox}, which we call MPOR-Fast, as an alternative to reduce the running time. In MPOR-Fast, we first solve the MPIC problem. If the embedding succeeds, we get the allocation $c^v$ and fix the corresponding path splitting ratios $\mathbf{f}^v(e^s)$. Next, $\forall e^s \in E^s$ such that $c^v(e^s) > 0$, we update $c^v(e^s)$ through LP in Equation (\ref{eq_ob_link_bandwidth}), which has only $O(N)$ decision variables and $O(M^v + N)$ constraints. This helps us save running time by solving no more than $|E^s|$ additional small LP problems, instead of enlarging the LP problem for MPIC.

\subsection{Multi-path Adaptive Routing (MPAR)} \label{ss_mcf_adaptive}

\begin{figure}[tbp]
\centering
\includegraphics[scale=0.8]{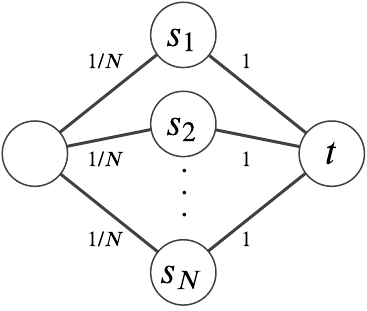}
\caption{An example of SN such that the $\mathcal{H}$-description and $\mathcal{V}$-description of $D^s$ are exponential in the description of SN.}
\label{fig_exp_face}
\end{figure}

For MPAR problem, its complexity remains unclear and it might be NP-hard. Suppose the feasible region of $\mathbf{d}^v$ with respect to the SN (including $G^s$ and $c^s$) is denoted by $D^s$, which is a polytope. Then the polytope containment problem of whether $D^v \subseteq D^s$ is equivalent to whether an MPAR algorithm will return a feasible $G^v$. It has been known that many forms of the polytope containment problem are polynomial \cite{freund1985complexity,kaibel2003some}, e.g., the two polytopes both have $\mathcal{H}$-descriptions (intersection of a finite set of half-spaces) or $\mathcal{V}$-descriptions (convex hull of a finite set of vertices). However, in our problem, either the $\mathcal{H}$-description or $\mathcal{V}$-description of $D^s$ can be exponential in the description of the SN. For example, for the SN shown in Fig. \ref{fig_exp_face} on which we specify $N$ node pairs $(s_1, t), (s_2, t), \cdots, (s_N, t)$, the associated half-spaces of $D^s$ are as follows:
\begin{align}
& \forall n \in \{1, 2, \cdots, N\}: \quad 0 \leq \mathbf{d}^v(n) \leq 1 + \frac{1}{N},  \label{eq_exp_face_1}\\
& \forall H \subseteq \{1, 2, \cdots, N\}\; s.t.\; |H| > \lfloor N / 2 \rfloor: \notag\\
& \quad \sum_{n\in H} \mathbf{d}^v(n) \leq |H| + \frac{N-|H|}{N}. \label{eq_exp_face_2}
\end{align}
In Equation (\ref{eq_exp_face_2}), for $H$'s such that $|H| = \lfloor N / 2 \rfloor + 1$ only, the number of half-spaces is $\binom N {\lfloor N / 2 \rfloor + 1}$, which is exponential in $N$. It can be similarly verified that the number of vertices of $D^s$ is also exponential in $N$. This means MPAR cannot be reduced to polytope containment problem efficiently. Moreover, reference \cite{freund1985complexity} has proved that the problem of whether a polytope with $\mathcal{H}$-description is contained in a polytope with $\mathcal{V}$-description is co-NP-complete. Therefore,  even if we focus on a subset of the MPAR problem where the $\mathcal{V}$-description of $D^s$ is polynomial in the description of the SN (remember $D^v$ has $\mathcal{H}$-description in our problem formulation), it is still unclear whether this subset is polynomial or not.

If the scale of the considered MPAR problem is relatively small, we can formulate an LP problem with exponential number of constraints that is solvable in reasonable time.  
% Since generally $|E^s| >> M^v$, we formulate LP through the first way described above. 
We first enumerate all the vertices of $D^v$, which takes $O(PNM^v)$ time \cite{avis1992pivoting} where $P$ is the number of the vertices. We denote the vertices by $\mathbf{d}^v_1, \mathbf{d}^v_2, \cdots, \mathbf{d}^v_P$. Then we exclude those vertices dominated by at least one other vertex, where we say $\mathbf{d}^v_p$ is dominated by $\mathbf{d}^v_{p'} \neq \mathbf{d}^v_p$ if and only if $\mathbf{d}^v_{p'} - \mathbf{d}^v_p \geq \mathbf{0}$. This can be done in $O(P^2N)$ time through brute-force method, and without loss of generality we assume the remaining vertices are the first $Q$ vertices. Next, for each vertex, we define a set of flow variables through the same way in Section \ref{ss_mcf_independent}. However, since now we consider $Q$ vertices, while in Section \ref{ss_mcf_independent} we only consider one vertex $\mathbf{d}^v_\text{max}$, we need to make a slight modification of the notations: $\forall n \in \{1, 2, \cdots, N\}, \forall e^s \in E^s$, instead of defining flow variables $f^v_n(e^s_+)$ and $f^v_n(e^s_-)$, we define $f^v_{n,q}(e^s_+)$ and $f^v_{n,q}(e^s_-)$ for each vertex $\mathbf{d}^v_q$, $q \in \{1, 2,\cdots, Q\}$. Hence, the final LP would have an objective function in Equation (\ref{eq_objective}), a set of SN link bandwidth constraints in Equation (\ref{eq_ind_link_capacity}), and a set of flow conservation, flow non-negativity and traffic demand constraints in Equation (\ref{eq_ind_flow_conservation}), (\ref{eq_ind_flow_nonnegativity}) and (\ref{eq_ind_link_bandwidth}) for each vertex $\mathbf{d}^v_q$. 
% The input complexity of this LP problem is also summarized in Table \ref{tab_lp_complexity}.

% Table \ref{tab_lp_complexity} summarizes the LP input complexity for MPIC, MPOR and MPAR. 

\section{Evaluation} \label{s_evaluation}

In this section, we evaluate the performance of different VNE algorithms through simulation. We first describe the settings of our simulation in Section \ref{ss_settings}, and then present the results of in Section \ref{ss_results}. 

\subsection{Settings} \label{ss_settings}

\textbf{Simulation Environment.} We run VNE simulation on a laptop with 2.7 GHz Intel Core I5 processor and 8-GB 1867 MHz DDR3 memory. The simulator is implemented in MATLAB, and IBM ILOG CPLEX Optimization Studio \cite{CPLEX} is used for solving LP. 

\begin{figure}[tbp]
\centering
\includegraphics[scale=0.2]{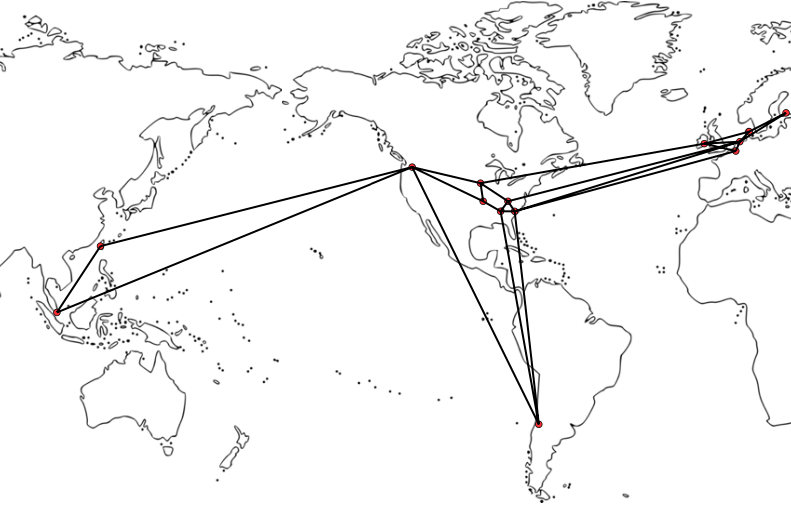}
\caption{B4 topology}
\label{fig_b4}
\end{figure}

\begin{figure*}[t]
\centering
\includegraphics[]{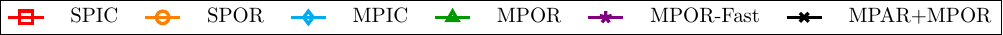}
\end{figure*}
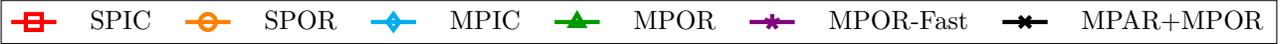
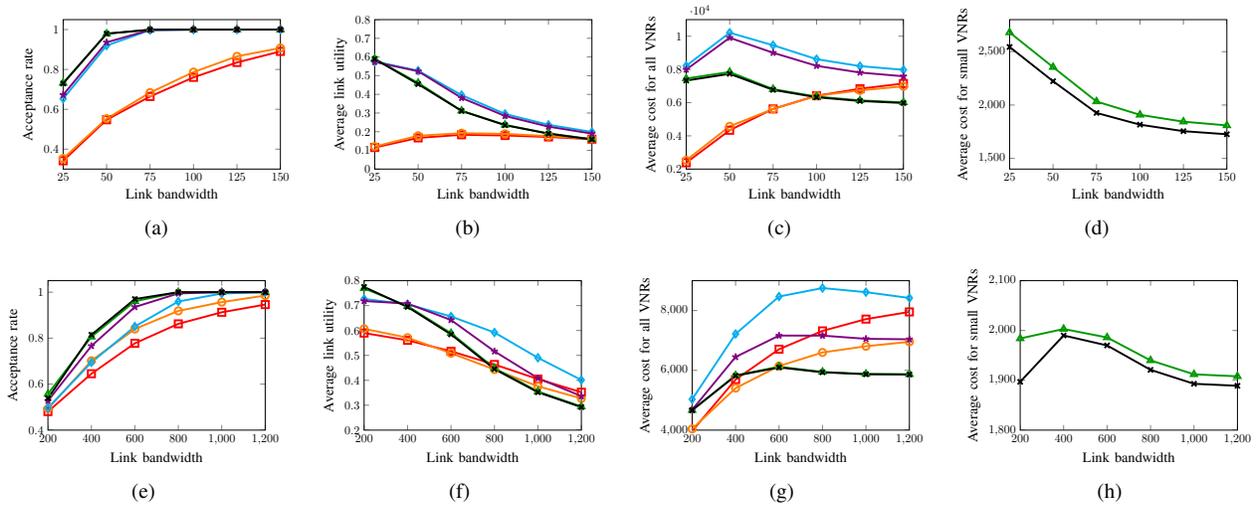
\begin{figure*}[t]
\centering
\subfigure[\label{sim_1_acceptance}]{
\begin{tikzpicture}[scale=0.45]
    \pgfplotsset{normalsize, samples=10}
    \begin{axis}[ height=6cm, width=8cm,
                  xmin=25,xmax=150,ymin=0.3,ymax=1.05,
                  xtick distance=25,
                  xlabel= Link bandwidth,
                  ylabel=Acceptance rate,
                  xlabel style={font=\large},
                  ylabel style={font=\large},
                  every axis plot/.append style={ultra thick},
                  mark size=3pt ]
            \addplot[red,mark=square] coordinates {
(25, 0.3416)
(50, 0.5474)
(75, 0.6642)
(100, 0.7602)
(125, 0.8350)
(150, 0.8905)
            };
            \addplot[orange,mark=o] coordinates {
(25, 0.3531)
(50, 0.5528)
(75, 0.6839)
(100, 0.7868)
(125, 0.8658)
(150, 0.9075)
            }; 
            \addplot[cyan,mark=diamond] coordinates {
(25, 0.6530)
(50, 0.9198)
(75, 0.9973)
(100, 1)
(125, 1)
(150, 1)
            };
            \addplot[green!60!black,mark=triangle] coordinates {
(25, 0.7321)
(50, 0.9792)
(75, 0.9992)
(100, 1)
(125, 1)
(150, 1)
            };
            \addplot[violet,mark=star] coordinates {
(25, 0.6723)
(50, 0.9360)
(75, 0.9981)
(100, 0.9996)
(125, 1)
(150, 1)
            };
            \addplot[black,mark=x] coordinates {
(25, 0.7298)
(50, 0.9796)
(75, 0.9996)
(100, 1)
(125, 1)
(150, 1)
            };
    \end{axis}
 
\end{tikzpicture}
}
\subfigure[\label{sim_1_util}]{
\begin{tikzpicture}[scale=0.45]
    \pgfplotsset{normalsize, samples=10}
    \begin{axis}[ height=6cm, width=8cm,
                  xmin=25,xmax=150,ymin=0,ymax=0.8,
                  xtick distance=25,
                  xlabel= Link bandwidth,
                  ytick distance=0.1,
                  ylabel=Average link utility,
                  xlabel style={font=\large},
                  ylabel style={font=\large},
                  every axis plot/.append style={ultra thick},
                  mark size=3pt ]
            \addplot[red,mark=square] coordinates {
(25, 0.1153)
(50, 0.1668)
(75, 0.1835)
(100, 0.1797)
(125, 0.1717)
(150, 0.1600)
            };
            \addplot[orange,mark=o] coordinates {
(25, 0.1206)
(50, 0.1783)
(75, 0.1922)
(100, 0.1885)
(125, 0.1762)
(150, 0.1626)
            }; 
            \addplot[cyan,mark=diamond] coordinates {
(25, 0.5760)
(50, 0.5267)
(75, 0.3960)
(100, 0.2952)
(125, 0.2371)
(150, 0.1987)
            };
            \addplot[green!60!black,mark=triangle] coordinates {
(25, 0.5903)
(50, 0.4612)
(75, 0.3123)
(100, 0.2364)
(125, 0.1908)
(150, 0.1599)
            };
            \addplot[violet,mark=star] coordinates {
(25, 0.5734)
(50, 0.5236)
(75, 0.3809)
(100, 0.2843)
(125, 0.2271)
(150, 0.1897)
            };
            \addplot[black,mark=x] coordinates {
(25, 0.5893)
(50, 0.4547)
(75, 0.3113)
(100, 0.2350)
(125, 0.1895)
(150, 0.1586)
            };
    \end{axis}
 
\end{tikzpicture}
}
\subfigure[\label{sim_1_price}]{
\begin{tikzpicture}[scale=0.45]
    \pgfplotsset{normalsize, samples=10}
    \begin{axis}[ height=6cm, width=8cm,
                  xmin=25,xmax=150,ymin=2000,ymax=11000,
                  xtick distance=25,
                  xlabel= Link bandwidth,
                  ylabel=Average cost for all VNRs,
                  xlabel style={font=\large},
                  ylabel style={font=\large},
                  every axis plot/.append style={ultra thick},
                  mark size=3pt ]
            \addplot[red,mark=square] coordinates {
(25, 2381)
(50, 4337)
(75, 5623)
(100, 6417)
(125, 6837)
(150, 7154)
            };
            \addplot[orange,mark=o] coordinates {
(25, 2526)
(50, 4574)
(75, 5616)
(100, 6413)
(125, 6736)
(150, 6984)
            }; 
            \addplot[cyan,mark=diamond] coordinates {
(25, 8218)
(50, 10211)
(75, 9457)
(100, 8617)
(125, 8200)
(150, 7976)
            };
            \addplot[green!60!black,mark=triangle] coordinates {
(25, 7454)
(50, 7845)
(75, 6815)
(100, 6366)
(125, 6138)
(150, 6014)
            };
            \addplot[violet,mark=star] coordinates {
(25, 7991)
(50, 9908)
(75, 9004)
(100, 8217)
(125, 7813)
(150, 7589)
            };
            \addplot[black,mark=x] coordinates {
(25, 7331)
(50, 7734)
(75, 6770)
(100, 6325)
(125, 6099)
(150, 5977)
            };
    \end{axis}
 
\end{tikzpicture}
}
\subfigure[\label{sim_1_price_small}]{
\begin{tikzpicture}[scale=0.45]
    \pgfplotsset{normalsize, samples=10}
    \begin{axis}[ height=6cm, width=8cm,
                  xmin=25,xmax=150,ymin=1400,ymax=2800,
                  xtick distance=25,
                  xlabel= Link bandwidth,
                  ylabel=Average cost for small VNRs,
                  xlabel style={font=\large},
                  ylabel style={font=\large},
                  every axis plot/.append style={ultra thick},
                  mark size=3pt ]
            \addplot[green!60!black,mark=triangle] coordinates {
(25, 2680)
(50, 2355)
(75, 2033)
(100, 1908)
(125, 1844)
(150, 1809)
            };
            \addplot[black,mark=x] coordinates {
(25, 2544)
(50, 2223)
(75, 1926)
(100, 1816)
(125, 1755)
(150, 1726)
            };
    \end{axis}
 
\end{tikzpicture}
}\\
\subfigure[\label{sim_2_acceptance}]{
\begin{tikzpicture}[scale=0.45]
    \pgfplotsset{normalsize, samples=10}
    \begin{axis}[ height=6cm, width=8cm,
                  xmin=200,xmax=1200,ymin=0.4,ymax=1.05,
                  xtick distance=200,
                  xlabel= Link bandwidth,
                  ylabel=Acceptance rate,
                  xlabel style={font=\large},
                  ylabel style={font=\large},
                  every axis plot/.append style={ultra thick},
                  mark size=3pt ]
            \addplot[red,mark=square] coordinates {
(200, 0.4807)
(400, 0.6453)
(600, 0.7776)
(800, 0.8622)
(1000, 0.9126)
(1200, 0.9461)
            };
            \addplot[orange,mark=o] coordinates {
(200, 0.4933)
(400, 0.7016)
(600, 0.8398)
(800, 0.9185)
(1000, 0.9567)
(1200, 0.9854)
            }; 
            \addplot[cyan,mark=diamond] coordinates {
(200, 0.4965)
(400, 0.6945)
(600, 0.8508)
(800, 0.9594)
(1000, 0.9941)
(1200, 1)
            };
            \addplot[green!60!black,mark=triangle] coordinates {
(200, 0.5575)
(400, 0.8059)
(600, 0.9603)
(800, 1)
(1000, 1)
(1200, 1)
            };
            \addplot[violet,mark=star] coordinates {
(200, 0.5276)
(400, 0.7661)
(600, 0.9354)
(800, 0.9945)
(1000, 1)
(1200, 1)
            };
            \addplot[black,mark=x] coordinates {
(200, 0.5382)
(400, 0.8146)
(600, 0.9701)
(800, 1)
(1000, 1)
(1200, 1)
            };
    \end{axis}
 
\end{tikzpicture}
}
\subfigure[\label{sim_2_util}]{
\begin{tikzpicture}[scale=0.45]
    \pgfplotsset{normalsize, samples=10}
    \begin{axis}[ height=6cm, width=8cm,
                  xmin=200,xmax=1200,ymin=0.2,ymax=0.8,
                  xtick distance=200,
                  xlabel= Link bandwidth,
                  ytick distance=0.1,
                  ylabel=Average link utility,
                  xlabel style={font=\large},
                  ylabel style={font=\large},
                  every axis plot/.append style={ultra thick},
                  mark size=3pt ]
            \addplot[red,mark=square] coordinates {
(200, 0.5897)
(400, 0.5606)
(600, 0.5166)
(800, 0.4635)
(1000, 0.4050)
(1200, 0.3521)
            };
            \addplot[orange,mark=o] coordinates {
(200, 0.6067)
(400, 0.5711)
(600, 0.5089)
(800, 0.4433)
(1000, 0.3761)
(1200, 0.3272)
            }; 
            \addplot[cyan,mark=diamond] coordinates {
(200, 0.7266)
(400, 0.7054)
(600, 0.6564)
(800, 0.5920)
(1000, 0.4905)
(1200, 0.4012)
            };
            \addplot[green!60!black,mark=triangle] coordinates {
(200, 0.7692)
(400, 0.6981)
(600, 0.5892)
(800, 0.4488)
(1000, 0.3552)
(1200, 0.2948)
            };
            \addplot[violet,mark=star] coordinates {
(200, 0.7183)
(400, 0.7080)
(600, 0.6422)
(800, 0.5164)
(1000, 0.4087)
(1200, 0.3364)
            };
            \addplot[black,mark=x] coordinates {
(200, 0.7750)
(400, 0.6945)
(600, 0.5843)
(800, 0.4450)
(1000, 0.3520)
(1200, 0.2922)
            };
    \end{axis}
 
\end{tikzpicture}
}
\subfigure[\label{sim_2_price}]{
\begin{tikzpicture}[scale=0.45]
    \pgfplotsset{normalsize, samples=10}
    \begin{axis}[ height=6cm, width=8cm,
                  xmin=200,xmax=1200,ymin=4000,ymax=9000,
                  xtick distance=200,
                  xlabel= Link bandwidth,
                  ylabel=Average cost for all VNRs,
                  xlabel style={font=\large},
                  ylabel style={font=\large},
                  every axis plot/.append style={ultra thick},
                  mark size=3pt ]
            \addplot[red,mark=square] coordinates {
(200, 3971)
(400, 5691)
(600, 6707)
(800, 7320)
(1000, 7714)
(1200, 7955)
            };
            \addplot[orange,mark=o] coordinates {
(200, 4042)
(400, 5406)
(600, 6144)
(800, 6598)
(1000, 6806)
(1200, 6954)
            }; 
            \addplot[cyan,mark=diamond] coordinates {
(200, 5033)
(400, 7218)
(600, 8470)
(800, 8753)
(1000, 8616)
(1200, 8421)
            };
            \addplot[green!60!black,mark=triangle] coordinates {
(200, 4668)
(400, 5832)
(600, 6121)
(800, 5948)
(1000, 5878)
(1200, 5870)
            };
            \addplot[violet,mark=star] coordinates {
(200, 4680)
(400, 6448)
(600, 7163)
(800, 7162)
(1000, 7055)
(1200, 7039)
            };
            \addplot[black,mark=x] coordinates {
(200, 4667)
(400, 5820)
(600, 6098)
(800, 5935)
(1000, 5870)
(1200, 5862)
            };
    \end{axis}
 
\end{tikzpicture}
}
\subfigure[\label{sim_2_price_small}]{
\begin{tikzpicture}[scale=0.45]
    \pgfplotsset{normalsize, samples=10}
    \begin{axis}[ height=6cm, width=8cm,
                  xmin=200,xmax=1200,ymin=1800,ymax=2100,
                  xtick distance=200,
                  xlabel= Link bandwidth,
                  ylabel=Average cost for small VNRs,
                  xlabel style={font=\large},
                  ylabel style={font=\large},
                  every axis plot/.append style={ultra thick},
                  mark size=3pt ]
            \addplot[green!60!black,mark=triangle] coordinates {
(200, 1984)
(400, 2003)
(600, 1986)
(800, 1940)
(1000, 1912)
(1200, 1908)
            };
            \addplot[black,mark=x] coordinates {
(200, 1897)
(400, 1990)
(600, 1970)
(800, 1921)
(1000, 1893)
(1200, 1889)
            };
    \end{axis}
 
\end{tikzpicture}
}
\caption{Embedding results for GT-ITM topology (top) and B4 topology (bottom).}
\label{fig_sims}
\end{figure*}

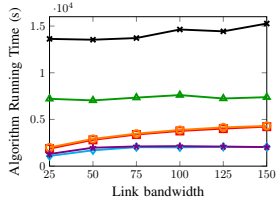
\begin{figure}[tbp]
\centering
\begin{tikzpicture}[scale=0.45]
    \pgfplotsset{normalsize, samples=10}
    \begin{axis}[ height=6cm, width=8cm,
                  xmin=25,xmax=150,ymin=0,ymax=16000,
                  xtick distance=25,
                  xlabel= Link bandwidth,
                  ylabel=Algorithm Running Time (s),
                  xlabel style={font=\large},
                  ylabel style={font=\large},
                  every axis plot/.append style={ultra thick},
                  mark size=3pt ]
            \addplot[red,mark=square] coordinates {
(25, 1873)
(50, 2815)
(75, 3386)
(100, 3767)
(125, 4018)
(150, 4230)
            };
            \addplot[orange,mark=o] coordinates {
(25, 2006)
(50, 2917)
(75, 3481)
(100, 3876)
(125, 4151)
(150, 4310)
            }; 
            \addplot[cyan,mark=diamond] coordinates {
(25, 1090)
(50, 1696)
(75, 2034)
(100, 2014)
(125, 2065)
(150, 2052)
            };
            \addplot[green!60!black,mark=triangle] coordinates {
(25, 7211)
(50, 7038)
(75, 7337)
(100, 7608)
(125, 7244)
(150, 7384)
            };
            \addplot[violet,mark=star] coordinates {
(25, 1301)
(50, 1974)
(75, 2112)
(100, 2140)
(125, 2095)
(150, 2038)
            };
            \addplot[black,mark=x] coordinates {
(25, 13625)
(50, 13535)
(75, 13715)
(100, 14631)
(125, 14433)
(150, 15267)
            };
    \end{axis}
 
\end{tikzpicture}
\caption{Algorithm running time for GT-ITM topology.}
\label{sim_1_time}
\end{figure}

\textbf{SN.} We run VNE simulation on two different SN topologies: 1) GT-ITM topology, a topology generated by GT-ITM tool \cite{zegura1996model}. In our simulation, the GT-ITM topology is configured to have 100 nodes which are randomly placed on a 100 $\times$ 100 grid, and each node pair is randomly connected with probability 0.1. So there are around 500 links in total. 2) B4 topology, the topology of Google's private inter-datacenter WAN \cite{jain2013b4, hong2018b4, GoogleDC}.
Fig. \ref{fig_b4} shows the B4 topology used in our simulation,
which has 14 nodes and 25 links. For each topology, the unit price of each link is proportional to the geographical distance between the two nodes at its two ends; the bandwidths of all the links are the same and we run simulation with six different link bandwidth values.

\textbf{VNR.} In each VNR, the number of access nodes follows a uniform distribution between 2 and 10. Each pair of access nodes is factored into the node pair vector $\mathbf{e}^v$ with probability 0.5. For each pair in $\mathbf{e}^v$, we specify a traffic demand upper bound that follows a uniform distribution between 1 and 20. Besides, we also add $N$ random joint traffic demand upper bound constraints, where each constraint specifies a joint traffic demand upper bound for two or more pairs of nodes which is smaller than the sum of their individual traffic demand upper bounds.

\textbf{Time Dimension.} The arrival of VNRs follows a Poisson process with an average rate of 5 VNRs per time unit, and the duration of each VNR follows an exponential distribution with an average of 10 time units. Each simulation lasts for 500 time units, which corresponds to around 2500 VNRs.

\textbf{VNE Algorithms.} We compare six VNE algorithms: SPIC, SPOR, MPIC, MPOR, MPOR-Fast, and MPAR+MPOR. In particular, for MPAR+MPOR, we run MPAR on small VNRs with $N \leq 5$ and MPOR otherwise, due to the time complexity of MPAR. 

\subsection{Results} \label{ss_results}

The embedding results for GT-ITM topology and B4 topology are shown in Fig. \ref{fig_sims}. The acceptance rate, average link utility, and average cost for all accepted VNRs and all accepted small VNRs of using different VNE algorithms are measured for both two topologies. 
Besides, we also present the algorithm running time of different algorithms for GT-ITM topology in Fig. \ref{sim_1_time}, since the scale of GT-ITM topology is large enough to make algorithm running time difference obvious. The key observations are summarized as follows.

\textbf{Multi-path embedding and shared channel embedding improve the acceptance rate when link bandwidth is limited.} Fig. \ref{sim_1_acceptance} and Fig. \ref{sim_2_acceptance} show the acceptance rate for GT-ITM topology and B4 topology, respectively. In both two figures, 
% we observe that, 
the performance of MPIC is always better than SPIC, and the performance of MPOR/MPOR-Fast is always better than SPOR, which indicates multi-path embedding improves acceptance rate. Moreover, the performance of SPOR is always better than SPIC, and when link bandwidth is smaller than 75 for GT-ITM and 1000 for B4 (i.e., when link bandwidth is limited for MPIC), the performance of MPOR/MPOR-Fast/MPAR+MPOR is better than MPIC, which indicates shared channel embedding improves acceptance rate. The reason is multi-path embedding and shared channel embedding require a cheaper allocation
% smaller SN subset 
and thus lead to higher acceptance rate.

\textbf{Shared channel embedding reduces link utility and cost when accepted VNRs are the same.} When link bandwidth is no less than 75 for GT-ITM and 1000 for B4, all the VNRs are accepted for MPIC, MPOR, MPOR-Fast, and MPAR+MPOR. Therefore, the corresponding link utility and cost are comparable. In Fig. \ref{sim_1_util}/Fig. \ref{sim_2_util} and Fig. \ref{sim_1_price}/Fig. \ref{sim_2_price}, in terms of average link utility and cost, we have MPAR+MPOR $<$ MPOR $<$ MPOR-Fast $<$ MPIC. The savings on cost for MPOR, MPOR-Fast, and MPAR+MPOR with respect to MPIC are $24.6\%$, $4.9\%$, and $25.1\%$ respectively for GT-ITM when link bandwidth is 150, and $30.3\%$, $16.4\%$, and $30.4\%$ respectively for B4 when link bandwidth is 1200. The reason is shared channel embedding requires a cheaper allocation
% smaller SN subset 
which reduces link utility and cost.

\textbf{For single-path embedding, the advantage of shared channel embedding is more significant in a smaller SN.} The advantage of SPOR, including acceptance rate, average link utility and average cost, is nearly negligible compared to SPIC for GT-ITM. However, for B4, we can see significant advantage of SPOR against SPIC in terms of acceptance rate and average cost. The main reason is the scale of B4 is relatively small, and therefore 
% for single-path embedding 
different node pairs have higher chances to use a common link.
 % which enhances the advantage of shared channel embedding.

\textbf{Compared to oblivious routing, adaptive routing has limited savings on cost for small VNRs.} Within all the VNRs, there are around 40\% small VNRs with $N \leq 5$. Fig. \ref{sim_1_price_small} and Fig. \ref{sim_2_price_small} show the average cost of these small VNRs for MPOR and MPAR (in MPAR+MPOR only MPAR is called for small VNRs). The saving on cost for MPAR with respect to MPOR is only $4.6\%$ for GT-ITM when link bandwidth is 150, and $1.0\%$ for B4 when link bandwidth is 1200. This implies that though adaptive routing tries to reduce the cost by offering more flexibilities, the associated advantage reflected on cost tends to be small.

\textbf{Shared channel embedding requires longer algorithm running time.} For GT-ITM topology, Fig. \ref{sim_1_time} indicates that shared channel embedding algorithms generally have longer algorithm running time. However, the additional running time seems to be trivial for SPOR (compared to SPIC) and MPOR-Fast (compared to MPIC), while MPOR and MPAR+MPOR require $3.6\times$ and $7.4\times$ running time of MPIC when link bandwidth is 150, respectively. So it is also a tradeoff between algorithm running time and VNE acceptance rate when link bandwidth is limited, or between algorithm running time and VNE cost/link utility when link bandwidth is adequate.

\section{Discussion} \label{s_discussion}

In previous sections we have studied VNE algorithms under the most basic configuration, where only traffic demand constraints are considered and each link has a fixed unit price. In this section, we briefly discuss the scenarios where node constraints and delay constraints are considered and link cost is a piecewise linear or quadratic function of link bandwidth.

\subsection{Node Constraints} \label{ss_node_constraints}

Node constraints generally include CPU capacity constraints and node location constraints. CPU capacity constraints dictate the CPU requirement for each virtual node, and node location constraints dictate the potential SN nodes that each virtual node can be mapped to.

When node constraints are introduced, a node mapping step is needed to map each virtual node to a fixed SN node, and then the aforementioned algorithms can be used to determine the final SN subset. The node mapping algorithm doesn't have to take traffic demand constraints into consideration. For example, the greedy node mapping algorithm in \cite{yu2008rethinking} only considers the remaining CPU capacity and outgoing link bandwidths for a potential SN node during node mapping, so in this case our description of traffic demand will not affect the node mapping at all. Even for those node mapping algorithms which do consider traffic demand constraints, we can only consider independent traffic demand constraints in this phase.

\subsection{Delay Constraints} \label{ss_delay_constraints}

Delay constraints dictate that maximum acceptable delay
for routing traffic of each node pair. When they are introduced, the latency of each SN link is also given and we need to consider the delay of one or more paths selected for each node pair. 

Suppose the latency of SN link $e^s$ is $\delta(e^s)$, and the maximum acceptable delay for node pair $\mathbf{e}^v(n)$ is $\Delta(\mathbf{e}^v(n))$. Then for single-path embedding, we only need to add one check after line 7 of Algorithm \ref{alg_ufp_independent}: if the sum of $\delta(e^s)$'s for those $e^s$'s belong to path $k$ is greater than $\Delta(\mathbf{e}^v(n))$, we shouldn't use this path. For multi-path embedding, if we only care about the average latency of all the selected paths, then we can simply add the following constraint to the LP given in Section \ref{ss_mcf_independent}: 
\begin{align}
& \forall n \in \{1, 2, \cdots, N\}: \notag\\
& \quad \sum_{e^s} [f^v_{n}(e^s_+) + f^v_{n}(e^s_-)]\cdot\delta(e^s)  \leq \Delta(\mathbf{e}^v(n))
\end{align}
However, if the delay of each selected path is required to be bounded by $\Delta(\mathbf{e}^v(n))$, then binary variables have to be introduced into the optimization, making it become an NP-time algorithm. Alternatively, we can use approximation algorithms in this scenario, but it is out of the scope of this paper so we stop our discussion here.

\subsection{Piecewise Linear or Quadratic Link Cost} \label{ss_link_price}

In the basic problem, we assume each link $e^s$ has a fixed unit price $p^s(e^s)$, i.e., link cost is a linear function of the its available bandwidth on the resulted SN subset. Link cost can also be modeled as a piecewise linear or quadratic function of its available bandwidth. In this case, for single-path embedding, Algorithm \ref{alg_ufp_independent} remains unchanged: we only need to use the piecewise linear or quadratic function to compute link cost, and we still use the same way to search the $K$-least cost paths in line 4. For multi-path embedding, the LP given in Section \ref{ss_mcf_independent} changes to mixed-integer linear programming (MILP) or quadratic programming (QP) when link cost is a piecewise linear or quadratic function, respectively. For piecewise linear link cost, which has multiple intervals, we need to introduce binary variables to determine which interval is link bandwidth located in; and for quadratic link cost, we can simply change the objective function in Equation (\ref{eq_objective}) to a sum of quadratic link cost functions.

\section{Conclusion} \label{s_conclusion}

In this paper, we presented a VNE problem where each VNR is directly described by the traffic demand between different access node pairs, without specifying a particular virtual network topology. This VNR formulation makes it possible for the customer to specify joint traffic demand constraints, which can possibly lead to a smaller and hence cheaper SN subset. 
Moreover, three different groups of VNE variants were considered. Simulations showed that shared channel embedding, as a new embedding variant under the proposed VNR formulation, improves the acceptance rate and reduces cost and link utility compared to traditional independent channel embedding.

There are several directions along which we can extend this study. First, the complexity of the MPAR embedding problem remains unclear and is therefore worthwhile to be studied. Another direction is to consider the constraints of dynamics of adaptive routing embedding, which imposes limitations on the real-time updates of routing and is a quite realistic concern. Finally, the joint traffic demand constraints can be factored into the phase of node mapping, which would further increase the search space for valid SN subset and possibly lead to a better result. 

\bibliographystyle{IEEEtran}
\bibliography{ref}

\begin{thebibliography}{10}
\providecommand{\url}[1]{#1}
\csname url@samestyle\endcsname
\providecommand{\newblock}{\relax}
\providecommand{\bibinfo}[2]{#2}
\providecommand{\BIBentrySTDinterwordspacing}{\spaceskip=0pt\relax}
\providecommand{\BIBentryALTinterwordstretchfactor}{4}
\providecommand{\BIBentryALTinterwordspacing}{\spaceskip=\fontdimen2\font plus
\BIBentryALTinterwordstretchfactor\fontdimen3\font minus
  \fontdimen4\font\relax}
\providecommand{\BIBforeignlanguage}[2]{{%
\expandafter\ifx\csname l@#1\endcsname\relax
\typeout{** WARNING: IEEEtran.bst: No hyphenation pattern has been}%
\typeout{** loaded for the language `#1'. Using the pattern for}%
\typeout{** the default language instead.}%
\else
\language=\csname l@#1\endcsname
\fi
#2}}
\providecommand{\BIBdecl}{\relax}
\BIBdecl

\bibitem{mijumbi2015network}
R.~Mijumbi, J.~Serrat, J.-L. Gorricho, N.~Bouten, F.~De~Turck, and R.~Boutaba,
  ``Network function virtualization: State-of-the-art and research
  challenges,'' \emph{IEEE Communications surveys \& tutorials}, vol.~18,
  no.~1, pp. 236--262, 2015.

\bibitem{han2015network}
B.~Han, V.~Gopalakrishnan, L.~Ji, and S.~Lee, ``Network function
  virtualization: Challenges and opportunities for innovations,'' \emph{IEEE
  Communications Magazine}, vol.~53, no.~2, pp. 90--97, 2015.

\bibitem{anderson2005overcoming}
T.~Anderson, L.~Peterson, S.~Shenker, and J.~Turner, ``Overcoming the internet
  impasse through virtualization,'' \emph{Computer}, vol.~38, no.~4, pp.
  34--41, 2005.

\bibitem{fischer2013virtual}
A.~Fischer, J.~F. Botero, M.~T. Beck, H.~De~Meer, and X.~Hesselbach, ``Virtual
  network embedding: A survey,'' \emph{IEEE Communications Surveys \&
  Tutorials}, vol.~15, no.~4, pp. 1888--1906, 2013.

\bibitem{yu2008rethinking}
M.~Yu, Y.~Yi, J.~Rexford, and M.~Chiang, ``Rethinking virtual network
  embedding: substrate support for path splitting and migration,'' \emph{ACM
  SIGCOMM Computer Communication Review}, vol.~38, no.~2, pp. 17--29, 2008.

\bibitem{chowdhury2009virtual}
N.~M.~K. Chowdhury, M.~R. Rahman, and R.~Boutaba, ``Virtual network embedding
  with coordinated node and link mapping,'' in \emph{IEEE INFOCOM 2009}.\hskip
  1em plus 0.5em minus 0.4em\relax IEEE, 2009, pp. 783--791.

\bibitem{chowdhury2011vineyard}
M.~Chowdhury, M.~R. Rahman, and R.~Boutaba, ``Vineyard: Virtual network
  embedding algorithms with coordinated node and link mapping,'' \emph{IEEE/ACM
  Transactions on networking}, vol.~20, no.~1, pp. 206--219, 2011.

\bibitem{cheng2011virtual}
X.~Cheng, S.~Su, Z.~Zhang, H.~Wang, F.~Yang, Y.~Luo, and J.~Wang, ``Virtual
  network embedding through topology-aware node ranking,'' \emph{ACM SIGCOMM
  Computer Communication Review}, vol.~41, no.~2, pp. 38--47, 2011.

\bibitem{rahman2013svne}
M.~R. Rahman and R.~Boutaba, ``Svne: Survivable virtual network embedding
  algorithms for network virtualization,'' \emph{IEEE Transactions on Network
  and Service Management}, vol.~10, no.~2, pp. 105--118, 2013.

\bibitem{lu2006efficient}
J.~Lu and J.~Turner, ``Efficient mapping of virtual networks onto a shared
  substrate,'' 2006.

\bibitem{yao2018novel}
H.~Yao, X.~Chen, M.~Li, P.~Zhang, and L.~Wang, ``A novel reinforcement learning
  algorithm for virtual network embedding,'' \emph{Neurocomputing}, vol. 284,
  pp. 1--9, 2018.

\bibitem{yan2020automatic}
Z.~Yan, J.~Ge, Y.~Wu, L.~Li, and T.~Li, ``Automatic virtual network embedding:
  A deep reinforcement learning approach with graph convolutional networks,''
  \emph{IEEE Journal on Selected Areas in Communications}, vol.~38, no.~6, pp.
  1040--1057, 2020.

\bibitem{zhang2023multi}
P.~Zhang, N.~Chen, S.~Li, K.-K.~R. Choo, C.~Jiang, and S.~Wu, ``Multi-domain
  virtual network embedding algorithm based on horizontal federated learning,''
  \emph{IEEE Transactions on Information Forensics and Security}, 2023.

\bibitem{azar2004optimal}
Y.~Azar, E.~Cohen, A.~Fiat, H.~Kaplan, and H.~R{\"a}cke, ``Optimal oblivious
  routing in polynomial time,'' \emph{Journal of Computer and System Sciences},
  vol.~69, no.~3, pp. 383--394, 2004.

\bibitem{kinsy2009application}
M.~A. Kinsy, M.~H. Cho, T.~Wen, E.~Suh, M.~Van~Dijk, and S.~Devadas,
  ``Application-aware deadlock-free oblivious routing,'' in \emph{Proceedings
  of the 36th annual international symposium on Computer architecture}, 2009,
  pp. 208--219.

\bibitem{kleinberg1996approximation}
J.~M. Kleinberg, ``Approximation algorithms for disjoint paths problems,''
  Ph.D. dissertation, Massachusetts Institute of Technology, 1996.

\bibitem{eppstein1998finding}
D.~Eppstein, ``Finding the k shortest paths,'' \emph{SIAM Journal on
  computing}, vol.~28, no.~2, pp. 652--673, 1998.

\bibitem{assad1978multicommodity}
A.~A. Assad, ``Multicommodity network flows-a survey,'' \emph{Networks},
  vol.~8, no.~1, pp. 37--91, 1978.

\bibitem{freund1985complexity}
R.~M. Freund and J.~B. Orlin, ``On the complexity of four polyhedral set
  containment problems,'' \emph{Mathematical programming}, vol.~33, no.~2, pp.
  139--145, 1985.

\bibitem{kaibel2003some}
V.~Kaibel and M.~E. Pfetsch, ``Some algorithmic problems in polytope theory,''
  in \emph{Algebra, geometry and software systems}.\hskip 1em plus 0.5em minus
  0.4em\relax Springer, 2003, pp. 23--47.

\bibitem{avis1992pivoting}
D.~Avis and K.~Fukuda, ``A pivoting algorithm for convex hulls and vertex
  enumeration of arrangements and polyhedra,'' \emph{Discrete \& Computational
  Geometry}, vol.~8, no.~3, pp. 295--313, 1992.

\bibitem{CPLEX}
\BIBentryALTinterwordspacing
``Ibm ilog cplex optimization studio,'' 2020. [Online]. Available:
  \url{https://www.ibm.com/products/ilog-cplex-optimization-studio}
\BIBentrySTDinterwordspacing

\bibitem{zegura1996model}
E.~W. Zegura, K.~L. Calvert, and S.~Bhattacharjee, ``How to model an
  internetwork,'' in \emph{Proceedings of IEEE INFOCOM'96. Conference on
  Computer Communications}, vol.~2.\hskip 1em plus 0.5em minus 0.4em\relax
  IEEE, 1996, pp. 594--602.

\bibitem{jain2013b4}
S.~Jain, A.~Kumar, S.~Mandal, J.~Ong, L.~Poutievski, A.~Singh, S.~Venkata,
  J.~Wanderer, J.~Zhou, M.~Zhu \emph{et~al.}, ``B4: Experience with a
  globally-deployed software defined wan,'' in \emph{ACM SIGCOMM Computer
  Communication Review}, vol.~43, no.~4.\hskip 1em plus 0.5em minus 0.4em\relax
  ACM, 2013, pp. 3--14.

\bibitem{hong2018b4}
C.-Y. Hong, S.~Mandal, M.~Al-Fares, M.~Zhu, R.~Alimi, C.~Bhagat, S.~Jain,
  J.~Kaimal, S.~Liang, K.~Mendelev \emph{et~al.}, ``B4 and after: managing
  hierarchy, partitioning, and asymmetry for availability and scale in google's
  software-defined wan,'' in \emph{Proceedings of the 2018 Conference of the
  ACM Special Interest Group on Data Communication}.\hskip 1em plus 0.5em minus
  0.4em\relax ACM, 2018, pp. 74--87.

\bibitem{GoogleDC}
\BIBentryALTinterwordspacing
``Google data centers,'' 2020. [Online]. Available:
  \url{https://www.google.com/about/datacenters/locations/}
\BIBentrySTDinterwordspacing

\end{thebibliography}

\end{document}